\documentclass[preprint,showpacs,preprintnumbers,amsmath,amssymb]{revtex4}

\usepackage{graphicx}
\usepackage{dcolumn}
\usepackage{bm}

\begin{document}

\preprint{APS/123-QED}

\title{
$^{138}$La-$^{138}$Ce-$^{136}$Ce nuclear cosmochronometer of supernova neutrino process}

\author{T.~Hayakawa$^{1,2,*}$,  T.~Shizuma$^1$, T.~Kajino$^{2,3}$, K.~Ogawa$^{4}$, H.~Nakada$^{5}$}
\affiliation{
$^1$Kansai Photon Science Institute, Japan Atomic Energy Agency, Kizu, Kyoto 619-0215, Japan.\\
$^2$National Astronomical Observatory, Osawa, Mitaka, Tokyo 181-8588, Japan.\\
$^3$Department of Astronomy, School of Science, University of Tokyo, Tokyo 113-0033, Japan.\\
$^4$RIKEN, Nishina center, Hirosawa, Wako, Saitama 351-0198, Japan.\\
$^5$Department of Physics, Graduate School of Science, Chiba University, Inage, Chiba 263-8522, Japan.\\}
\date{\today}

\email[E-mail: ]{hayakawa.takehito@jaea.go.jp}

\begin{abstract}

The $^{138}$La (T$_{1/2}$=102 Gyr) - $^{138}$Ce - $^{136}$Ce system 
is proposed to be used as a nuclear cosmochronometer 
for measuring the time elapsed from a supernova neutrino process.
This chronometer is applied to examine a sample
affected by a single nucleosynthesis episode as presolar grains in primitive meteorites.
A feature of this chronometer is to evaluate the initial abundance ratio of  $^{136}$Ce/$^{138}$Ce
using an empirical scaling law, which was found in the solar abundances.
We calculate the age of the sample as a function of isotopic ratios, $^{136}$Ce/$^{138}$Ce,
and $^{138}$La/$^{138}$Ce,
and evaluate the age uncertainty due to theoretical and observational errors.
It is concluded that this chronometer can work well for
a sample with the abundance ratio of $^{138}$La/$^{138}$Ce $\ge$ 20
when the ratios of $^{136}$Ce/$^{138}$Ce and $^{138}$La/$^{138}$Ce are measured within the uncertainty of 20\%.
The availability of such samples becomes clear in recent studies of the presolar grains.
We also discuss the effect of the nuclear structure to the $\nu$ process origin of $^{138}$La.
\end{abstract}

\maketitle

Long-lived radioactivities  are used as nuclear cosmochronometers for dating
the ages of nucleosynthesis episodes, metal-poor stars \citep{Ur,Th}, and the Milky way \citep{Dauphas05}.
Rutherford suggested the original idea, the uranium chronometer, in 1929 \cite{Rutherford}.
Recent progress of astronomical observations 
has enabled to detect actinoid elements of early generations of stars \cite{Ur,Th}.
A fraction of the uranium in a metal-deficient star CS 31082-001 was measured
and the stellar age was estimated with the $^{238}$U chronometer for the first time in 2001 \cite{Ur}.
Only six chronometers with half-lives of 1-100 Gyr which are suitable for measuring cosmological ages
were known.
The other chronometers are $^{40}$K and $^{232}$Th suggested by Burbidge {\it et al.} \cite{BBFH},
$^{87}$Rb and $^{187}$Re suggested by Clayton \cite{Clayton64}, and
$^{176}$Lu suggested by Audouze {\it et al.} \cite{Audouze72a}.
It should be noted that these chronometers measure the age of slow or rapid neutron-capture reaction nucleosynthesis 
($s$ or $r$ process).
However, there is no cosmochronometer for neutrino-induced reactions in supernova explosions ($\nu$ process).

The $\nu$ process was suggested as the origin 
of heavy and light elements \cite{Woosley90,Goriely01,Rauscher02,Heger05,Yoshida05,Pruet06,Frohlich06,Yoshida06},
and is of importance for studying neutrino spectra from the supernovae \cite{Heger05,Yoshida05}
and for discussing neutrino oscillation \cite{Yoshida06}.
Among many heavy elements, only two isotopes of $^{138}$La and $^{180}$Ta
are considered to be synthesized primarily 
by the $\nu$ process \cite{Woosley90,Heger05},
but there is another possibility for $^{180}$Ta that is
produced by the $s$ process \cite{Yokoi83}.
The solar abundance of $^{138}$La
was quantitatively explained by the $\nu$ process in theoretical calculations \citep{Heger05,Byelikov07}.
Thus, $^{138}$La is a key for understanding the $\nu$ process.
In this paper, we propose
a new cosmochronometer based on the abundance ratios of the three isotopes,
$^{138}$La, $^{138}$Ce and $^{136}$Ce, in a single sample
affected strongly by a single nucleosynthesis episode.
Recent studies of presolar grains in primitive meteorites provide
samples affected strongly by single supernova nucleosynthesis episodes \citep{Amari92,Nittler03}.
The isotopic patterns of heavy elements such as molybdenum and barium
in the silicon-carbide (SiC) presolar grains originating from the supernovae have been measured \citep{Pellin06,Savina07}.
On the basis of these technological developments,
the abundances of $^{136,138}$Ce and $^{138}$La 
in a single sample as the presolar grain are expected to be determined in future.

The meta-stable isotope $^{138}$La
decays to $^{138}$Ce or $^{138}$Ba with a half-life of 102 Gyr.
The $^{138}$La-$^{138}$Ba (or $^{138}$Ce) 
system has been used as a geochronometer to date the ages of rocks \citep{Tanaka82}.
Audouze {\it et al.} pointed out that the $^{138}$La system 
is not useful as the cosmochronometer
since the abundance of the daughter nucleus populated by the decay of $^{138}$La may 
be miner relative to its initial abundance \citep{Audouze72a}.
However, nucleosynthesis models cannot predict generally the initial abundances
of individual presolar grains,
because the supernovae produce 
both of $^{138}$La and $^{138}$Ce and their initial abundance ratios
in individual grains are different.
After the destroy of pre-existing $^{138}$La and $^{138}$Ce (and $^{136}$Ce) 
by the weak $s$ process in evolutionary stages of progenitors,
these nuclei are re-synthesized in different layers in the supernova explosion;
$^{138}$La is synthesized by the $\nu$ process 
in He- and C-rich layers \citep{Heger05}
and $^{136,138}$Ce are synthesized 
by the $\gamma$ process in O/Ne layers \citep{Woosley78,Arnould03}.
Meteorite compositions are produced in stellar outflows
after mixture of materials originating from different layers,
which may include both the Ce isotopes and $^{138}$La.
The mixing ratios of the different layers in the individual meteorite compositions are different.
Therefore we should evaluate the initial abundances in each sample by an alternative method.

Here we propose a novel empirical method to estimate the initial abundance of $^{138}$Ce.
A similar empirical method was suggested for a $^{146}$Sm nuclear cosmochronometer,
wherein an initial abundance of $^{146}$Sm
is evaluated by the solar abundance pattern \citep{Audouze72b}.
In our proposed method, the initial abundance of $^{138}$Ce can be calculated 
by an empirical scaling law for $p$-nuclei, 
which was found in the solar abundances \citep{Hayakawa04}.
The three isotopes, $^{138}$La, $^{138}$Ce, and $^{136}$Ce, are 
classified to the $p$-nuclei.
The $p$-nuclei are located on the neutron-deficient
side from the $\beta$-stability line and their isotopic fractions are small (typically 0.1-1\%).
There are nine pairs of the two $p$-nuclei with the same atomic number.
As shown in Fig.~1, the first and second $p$-nuclei 
are lighter than the $s$-nucleus by two and four neutrons, respectively.
We found the empirical scalings that
the abundance ratios of $N$$_{\odot}$(1st $p$)/$N$$_{\odot}$(2nd $p$)
and $N$$_{\odot}$($s$)/$N$$_{\odot}$(1st $p$)
are almost constant over a wide range of atomic number, respectively,
where $N$$_{\odot}$ is the solar isotope abundance \citep{Hayakawa04}.
These scalings are a piece of evidence that the most probable origin of the $p$-nuclei
is the supernova $\gamma$ process.
In addition, we found a novel concept of "the universality of the $\gamma$ process", 
wherein the scalings hold for nuclides produced in individual supernova $\gamma$ processes
under various astrophysical conditions \citep{Hayakawa04}.
We calculated the $\gamma$ process of core-collapse supernovae under various 
astrophysical conditions 
such as, metallicity, progenitor mass, and explosion energy 
and found that the scalings hold for individual nucleosyntheses 
independent of the astrophysical conditions assumed \citep{Hayakawa06}.
We presented a detailed mechanism why the scalings fold for various supernovae in our previous papers \citep{Hayakawa06,Hayakawa08a}.
Our proposed chronometer is based on the universality of the scaling 
for the two $p$-nuclei.

Here we improve the empirical scaling for the two $p$-nuclei.
There are nine pairs of the two $p$-nuclei in the solar system,
but we omit three elements of Mo, Ru and Er in Fig.~\ref{fig:ratio}
because the origin of $p$-nuclei of Mo and Ru is considered to be different from
other $p$-nuclei \citep{Hayakawa04,Pruet06,Frohlich06}
and a $p$-nucleus $^{164}$Er is largely contributed from the $s$ process \citep{Jung92}.
The $N_{\odot}$(1st $p$)/$N_{\odot}$(2nd $p$) ratios
in the other six elements
increase with increasing the atomic number (see Fig.~2).
We obtain a function,
\begin{equation}
R_{pp}(Z) = 0.0581{\times}Z - 2.18,
\label{Rpp}
\end{equation}
using a $\chi$-square fitting.
As shown in Fig.~\ref{fig:ratio},
the observed $N_{\odot}$(1st $p$)/$N_{\odot}$(2nd $p$) ratios
are described well by this equation
and the standard deviation of $N_{\odot}$(1st $p$)/$N_{\odot}$(2nd $p$) is only 9\%.

If abundance ratios of N($^{138}$La)/N($^{138}$Ce) and N($^{136}$Ce)/N($^{138}$Ce) 
in a sample affected strongly by a single supernova
are known, the age $T$ elapsed from the supernova episode is calculated by
\[
T =
-\frac{T_{1/2}(^{138}\mbox{La})}{ln2}{\times}
\]
\begin{equation}
ln\Biggl(\frac{N(^{138}\mbox{La})/N(^{138}\mbox{Ce})}
{N(^{138}\mbox{La})/N(^{138}\mbox{Ce})+\frac{1}{b}
\Bigl(1-R_{pp}(\mbox{Ce}){\times}N(^{136}\mbox{Ce})/N(^{138}\mbox{Ce})\Bigr)}\
\Biggr), \label{chrono2}
\end{equation}
where $N$ is the observed abundance in the sample, $R_{pp}$ is the initial $N$(1st $p$)/$N$(2nd $p$) ratio
at the freezeout of the $\gamma$ process,
and $b$ is the branching ratio of the ${\beta}^{-}$ decay of $^{138}$La: 
${\lambda}({\beta}^{-})$ / (${\lambda}({\beta}^{-}$)+${\lambda}$(EC)).
The scaling holds for nuclides produced by individual supernova nucleosyntheses
and  the function (\ref{Rpp}) would hold for individual presolar grains.
The $R_{pp}$(Ce) ratio is estimated to be 1.19 using the function (\ref{Rpp}).
Therefore, the age can be obtained from the two observed ratios of 
N($^{138}$La)/N($^{138}$Ce) and N($^{136}$Ce)/N($^{138}$Ce)
in the sample.

A part of $^{138}$La may be contributed from the $\gamma$ process,
but the age obtained from this chronometer is not affected by the contamination
of the $\gamma$ process,
because this chronometer measure the age elapsed from a supernova 
and both the $\gamma$- and $\nu$ processes occur at the same supernova.
The cosmic ray process \citep{Audouze70}  
also has been suggested as the origin of $^{138}$La.
However, the presolar grains originating from the supernovae do not contain the comic ray products,
because the pre-existing $^{138}$La and $^{136,138}$Ce originating from the interstellar media,
which include the cosmic ray products,
are destroyed by the weak $s$ process before the supernovae.
The $p$-nuclei that include the three isotopes 
are located on the neutron-deficient side from the $\beta$-stability line, 
and hence they cannot 
be synthesized by neutron capture reactions in the weak $s$ process. 
On the other hand, the pre-existing $p$-nuclei
are destroyed by the neutron capture reactions.
Therefore, the three isotopes in the presolar grains from the supernovae
are newly synthesized by single nucleosyntheses without any pre-existing contamination
such as the cosmic ray products.

Here we discuss the uncertainty of the age using the chronometer.
Figure 3 shows the calculated ages as a function 
of $R_{pp}{\cdot}$N($^{136}$Ce)/N($^{138}$Ce)
under the mixing ratio N($^{138}$La)/N($^{138}$Ce) = 1, 5, 10 and 20.
We also calculated the ages as a function of N($^{138}$La)/N($^{138}$Ce)
under the ratio N($^{136}$Ce)/N($^{138}$Ce) = 1.5, 2.0, and 3.0 (see Fig.~4).
The uncertainty of the age 
is generally due to theoretical and observational errors.
In the present proposed chronometer, the theoretical errors come only from
the uncertainty of the initial abundance ratio of $R_{pp}$(Ce).
The uncertainty of $R_{pp}$(Ce) is estimated 
from the deviation of the observed $N_{\odot}$(1st p)/$N_{\odot}$(2nd p) ratios (see Fig.~2).
The standard deviation of the observed ratios is 9\% 
and we take 9\% as the uncertainty of $R_{pp}$(Ce).
The observational errors are due to
the errors of the N($^{136}$Ce)/N($^{138}$Ce)
and N($^{138}$La)/N($^{138}$Ce) ratios in samples.
Recently, measurement techniques with mass separation
have been rapidly developed
and the isotopic fractions of heavy elements such as Mo and Ba in SiC grains 
originating from the supernovae have been measured within the uncertainties of 
7-15\% \citep{Pellin06,Savina07}.
Thus, we consider two cases that the observational errors of both the two ratios
are 10 or 20\%.
Finally, we calculate the complete uncertainty
due to both the theoretical and observational errors; namely, the errors of
$R_{pp}$, N($^{136}$Ce)/N($^{138}$Ce), and N($^{138}$La)/N($^{138}$Ce).
If the ratios of  N($^{136}$Ce)/N($^{138}$Ce)
and N($^{138}$La)/N($^{138}$Ce) are measured with the uncertainty of 10\%, respectively,
the complete uncertainty is about 20\%, 50\% or 90\%
in the case of N($^{138}$La)/N($^{138}$Ce) = 20, 10 or 5.
With observational errors of 20\%,
the complete uncertainty turns out to be about 30\% for a sample of N($^{138}$La)/N($^{138}$Ce) = 20.
It should be noted that the complete uncertainty suddenly decreases with increasing the ratio of N($^{138}$La)/N($^{138}$Ce).

Recent progress of the astronomical observation has enabled to detect actinoid elements
of metal-poor stars.
The ages of these metal-poor stars have been evaluated with U and Th chronometers
within uncertainty of $\sim$50\% \citep{Ur,Th}. 
Therefore the present proposed chronometer 
is as useful as the cosmochronometers
have been used in the astronomy.

The discussion mentioned above shows that
the present proposed chronometer can work well if the presolar grains of 
N($^{138}$La)/N($^{138}$Ce) $\geq$ 20 are found with the observational errors of 20\%.
A question we have to ask here is whether the presolar grains of N($^{138}$La)/N($^{138}$Ce) $\geq$ 20
exist or not in nature.
Since the solar abundance ratio $N_{\odot}$($^{138}$La)/$N_{\odot}$($^{138}$Ce) is about 0.14,
it seems to be difficult to find the samples of N($^{138}$La)/N($^{138}$Ce) $\geq$ 20.
However, recent studies of the presolar grains originating from the supernovae 
suggest that such samples are available.
The carbon-rich grains of the SiC grains and low-density graphite grains
are chemically condensed in a carbon-rich environment of C/O $>$ 1.
This suggests that the main component of the carbon-rich grains
originating from the supernovae are produced in their C-rich layers, where $^{138}$La
is synthesized by the $\nu$ process.
In contrast, the contribution of O/Ne layers, where $^{136,138}$Ce 
are produced by the $\gamma$ process, is considered to be small.
The isotopic fractions of several elements in individual presolar grains
from supernovae were measured
and theoretical calculations were performed to reproduce
these isotopic fractions by proper mixture of 
materials produced in different layers \citep{Travaglio99,Yoshida04}.
The observed isotopic fractions were reproduced consistently 
without the contribution from the O/Ne layers \citep{Travaglio99}
and with the small contribution from the O/Ne layers which is smaller than that from the C-rich layers
by 2-4 orders of magnitude \citep{Yoshida04}. 
We stress again that $^{136,138}$Ce are dominantly synthesized in the O/Ne layers by the $\gamma$ process.
Therefore, we expect that samples with the abundance ratio of N($^{138}$La)/N($^{138}$Ce) $\geq$ 20
will be found.
Finally, we would like to stress the complete uncertainty suddenly decreases with increasing the ratio of N($^{138}$La)/N($^{138}$Ce).
The discussion mentioned above shows that we can find samples, of which the N($^{138}$La)/N($^{138}$Ce) ratios are much larger than 20,
and thereby the complete uncertainties of the ages of the sample become much lower than 30\%.

Although recent studies of the astrophysics suggest that $^{138}$La 
may be dominantly synthesized by the $\nu$ process \citep{Heger05,Byelikov07},
the origin of $^{138}$La has not been elucidated in astronomical observations
or meteorite analyses.
To take a hypothetical example, suppose that $^{138}$La is dominantly synthesized by the cosmic ray process
and the presolar grains are formed in an interstellar environment.
The age estimated by this chronometer would be longer than 
that in the case that $^{138}$La is dominantly synthesized by the $\nu$ process
because the cosmic ray nucleosynthesis is a continuing process from
the Galaxy formation to the solar system formation.
Thus, the age measured in the presolar grains
is of importance to understand the origin of $^{138}$La.

Here we point out that the nuclear structure of $^{138}$La
may affect the performance of the chronometer and the $\nu$ process origin of $^{138}$La.
A fact that the spin and parity of the ground state of an isotone $^{140}$Pr ($N$=81) is $J^{\pi}$=1$^+$
suggests that a 1$^+$ excited state may exist at a low excitation energy in $^{138}$La.
Although a 1$^+$ state at 296 keV in $^{138}$La was reported \cite{Islam},
the lowest 1$^+$ state has not been established experimentally.
If this 1$^+$ excited state exists at energy lower than 72 keV
that is the energy of the first excited state,
the 1$^+$ state may be a $\beta$ unstable isomer
 since $\beta$ decay
can compete with an internal $E4$ transition.
In such a case, our proposed clock cannot work well since
the initial abundance cannot be estimated by using the scaling.
In addition, this isomer affects the $\nu$ process origin of $^{138}$La
because $^{138}$La synthesized by the $\nu$ process is destroyed via the isomer (see Fig.~\ref{fig:1+}).

We would like to stress that the observed 1$^+$ ground state of $^{140}$Pr 
can not be understood by a simple shell-model picture that
would lead to the ground state with $J^{\pi}$=2$^+$ or 3$^+$ \cite{Helton73}.
This indicates that there is a mechanism decreasing the energy of the 1$^+$ state.
In order to examine the existence of the 1$^+$ isomer,
we calculate the nuclear structures of $^{138}$La and $^{140}$Pr
in a shell-model. 
In the mass region around $^{138}$La, the last proton (neutron) 
can occupy either d$_{5/2}$ or g$_{7/2}$ (d$_{3/2}$)
orbit around the Fermi surface.
The 1$^+$ and 5$^+$ states in $^{138}$La and $^{140}$Pr 
can be understood generally by (${\pi}$d$_{5/2}$)$\otimes$(${\nu}$d$_{3/2}$)$^{-1}$ 
and (${\pi}$g$_{7/2}$)$\otimes$(${\nu}$d$_{3/2}$)$^{-1}$ configurations, respectively,
where $\pi$ ($\nu$) is proton (neutron).
Figure \ref{fig:shell_model} show the calculated results.
We successfully describe the observed excited states 
in $^{138}$La \cite{Lasijo72,Helton73} and $^{140}$Pr \cite{Hussein74} 
lower than 500 keV and,
in particular, the ground state spin of $^{140}$Pr.
We find that
the coupling of two- and four-particle excited configurations 
of $\pi$(g$_{7/2}$)$^{-2}$(d$_{5/2}$)$^3$ and $\pi$(g$_{7/2}$)$^{-4}$(d$_{5/2}$)$^5$
leads to the 1$^+$ ground state. 
A calculated 1$^+$ state in $^{138}$La locates at 169 keV.
This result suggests that the 1$^+$ state in $^{138}$La is not an isomer
but an experimental determination is desired.

In summary, we propose 
a novel $^{138}$La (T$_{1/2}$=102 Gyr) - $^{138}$Ce - $^{136}$Ce nuclear cosmochronometer
for measuring the time elapsed from the supernova $\nu$ process.
We introduce a novel method 
using the universality of the empirical scaling for $p$ nuclei
to evaluate the initial abundance of $^{138}$Ce.
The chronometer is applied to examine a sample
affected by single supernova nucleosyntheses as presolar grains in primitive meteorites.
Recent studies of the presolar grains provide isotopic fractions of several different elements 
of single samples.
The chronometer can work well for a sample, 
of which its abundance ratio of $^{138}$La/$^{138}$Ce is larger than 20,
when the ratios of $^{136}$Ce/$^{138}$Ce and $^{138}$La/$^{138}$Ce are measured within the uncertainty of 20\%.
The availability of such samples becomes clear in recent studies of the presolar grains.
With the chronometer, we can measure both the age of the supernova associated with neutrino-driven wind
and the isotopic abundance patterns of elements synthesized by the supernova.
We present that the energy of the lowest 1$^+$ state
may affect the chronometer performance and the $\nu$ process origin of $^{138}$La,
and its measurement is desired.
\\

We thank T. Komatsubara, K. Yamashita and M. Fujiwara for valuable discussion.
This work has been supported in part by Grants-in-Aid for Scientific
from the Japan Society for the Promotion of Science
Research (18340071).

\clearpage

\begin{figure}
\includegraphics[viewport=0mm 0mm 120mm 70mm, clip, scale=0.7]{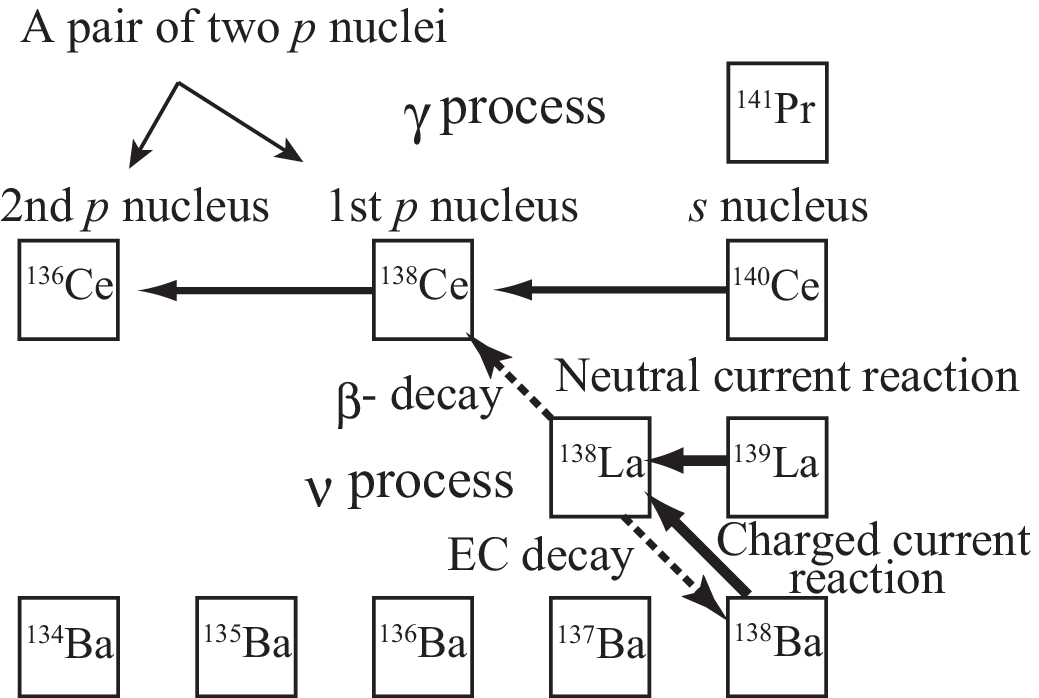}
\caption{
A partial nuclear chart and nucleosynthesis flows around $^{138}$La. 
The nucleus $^{138}$La is shielded by stable isobars against $\beta$ decay
and is produced by neither $s$ nor $r$ process.
In the $\nu$ process, $^{138}$La
is mainly synthesized from $^{138}$Ba via the charged current reaction.
Two $p$-nuclei $^{136}$Ce and $^{138}$Ce are mainly produced from 
an $s$-nucleus $^{140}$Ce via successive ($\gamma$,n) reactions.}
\label{fig:chart}
\end{figure}

\begin{figure}
\includegraphics[viewport=0mm 0mm 200mm 140mm, clip, scale=0.4]{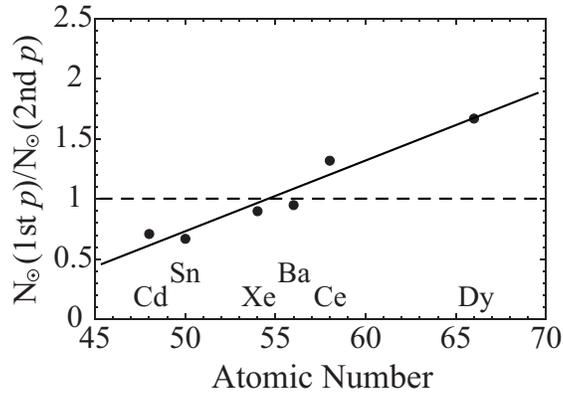}
\caption{
Observed isotope ratios of the first $p$-nuclei to the second $p$-nuclei with the same atomic number.
The solid line is the result of $\chi$-square fitting. The standard deviation
of the fitting is 9\%.
}
\label{fig:ratio}
\end{figure}

\begin{figure}
\includegraphics[viewport=0mm 0mm 200mm 140mm, clip, scale=0.4]{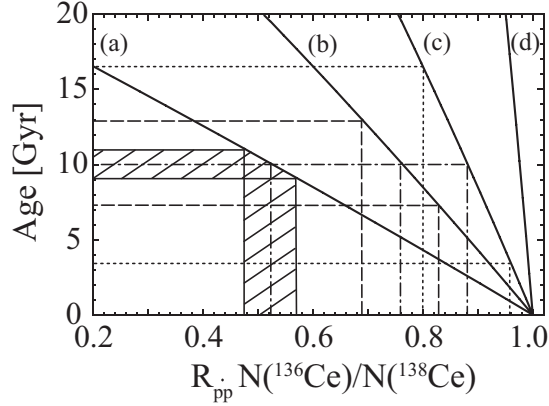}
\caption{
Calculated age as a function $R_{pp}$${\cdot}$N($^{136}$Ce)/N($^{138}$Ce).
The age is the time elapsed from a nucleosynthesis episode to the present,
which includes the solar system age 4.6 Gyr.
The solid curves of (a), (b), (c) and (d) are the calculated ages with
mixing ratios N($^{138}$La)/N($^{138}$Ce) = 20, 10, 5 and 1, respectively.
The dot-dashed line is the age of $T$=10 Gyr.
The solid, dashed, and dot lines
are the upper and lower limits arisen from uncertainty of $R_{pp}$
or N($^{136}$Ce)/N($^{138}$Ce)
for (a), (b) and (c), respectively. 
}
\label{fig:age}
\end{figure}

\begin{figure}
\includegraphics[viewport=0mm 0mm 200mm 140mm, clip, scale=0.4]{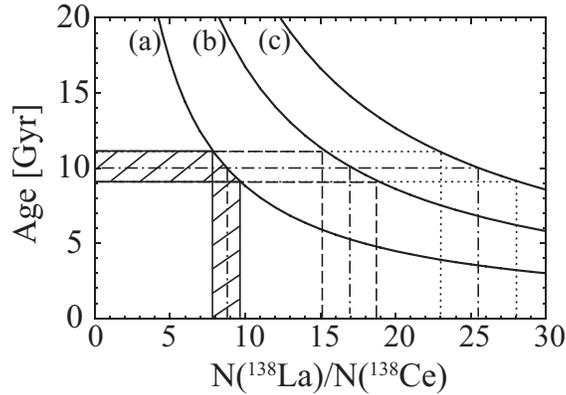}
\caption{
Calculated age as a function N($^{138}$La)/N($^{138}$Ce).
The age is the time elapsed from a nucleosynthesis episode.
The solid curves of (a), (b) and (c) are the calculated ages with
the ratios N($^{138}$Ce)/N($^{136}$Ce) = 1.5, 2.0 and 3.0, respectively.
The dot-dashed line is the age of $T$=10 Gyr.
The solid, dashed, and dot lines
are the upper and lower limits of the assumed uncertainty of N($^{138}$La)/N($^{138}$Ce)
for (a), (b) and (c), respectively. 
}
\label{fig:age_la}
\end{figure}

\begin{figure}
\includegraphics[viewport=0mm 0mm 200mm 140mm, clip, scale=0.6]{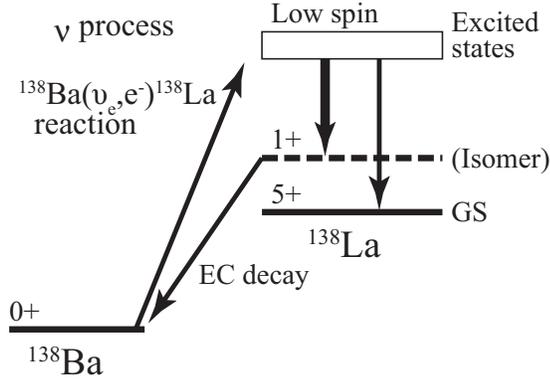}
\caption{
Synthesis and destruction of $^{138}$La in the $\nu$ process. Low spin excited states
such as $J^{\pi}$=1$^+$
are strongly populated from $^{138}$Ba by Gamow-Teller transitions.
If the 1$^+$ isomer exists,
these excited states strongly decay to the 1$^+$ isomer
which decays to daughter nuclei.
}
\label{fig:1+}
\end{figure}

\begin{figure}
\includegraphics[viewport=0mm 0mm 210mm 235mm, clip, scale=0.35]{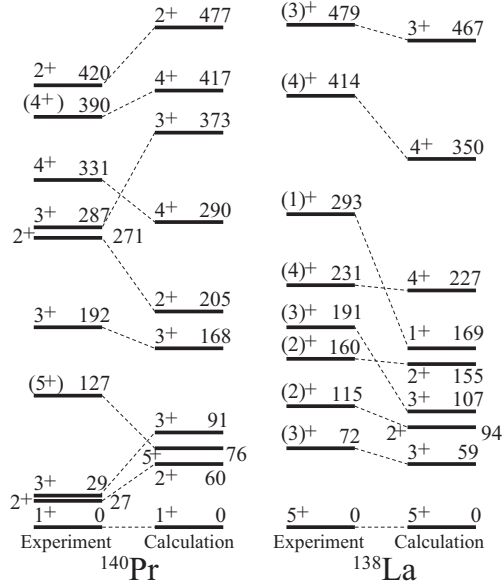}
\caption{
The calculated and measured levels in $^{140}$Pr and $^{138}$La.
The observed levels of $^{140}$Pr and $^{138}$La are taken from 
ref.~\cite{Hussein74} and refs.~\cite{Lasijo72,Helton73}, respectively.
}
\label{fig:shell_model}
\end{figure}

\end{document}